\documentclass[aps,floats]{revtex4}
\usepackage{amsmath,amssymb}
\usepackage{graphicx,epsfig}
\usepackage[greek,english]{babel}

\begin{document}
\bibliographystyle {plain}

\def\oppropto{\mathop{\propto}} 
\def\opsimeq{\mathop{\simeq}}
\def\opoverderline{\mathop{\overline}}
\def\operarrow{\mathop{\longrightarrow}}
\def\opsim{\mathop{\sim}} 
\def\opmin{\mathop{\min}} 
\def\opmax{\mathop{\max}} 

\def\fig#1#2{\includegraphics[height=#1]{#2}}
\def\figx#1#2{\includegraphics[width=#1]{#2}}


\title{ Finite size scaling for the Many-Body-Localization Transition :  \\
finite-size-pseudo-critical points of individual eigenstates } 


\author{ C\'ecile Monthus }
 \affiliation{Institut de Physique Th\'{e}orique, 
Universit\'e Paris Saclay, CNRS, CEA,
91191 Gif-sur-Yvette, France}

\begin{abstract}
To understand the finite-size-scaling properties of phases transitions in classical and quantum models in the presence of quenched disorder, it has proven to be fruitful to introduce the notion of a finite-size-pseudo-critical point in each disordered sample and to analyze its sample-to-sample fluctuations as a function of the size. For the Many-Body-Localization transition, where very strong eigenstate-to-eigenstate fluctuations have been numerically reported even within a given disordered sample at a given energy density [X. Yu, D. J. Luitz, B. K. Clark, arxiv:1606.01260 and V. Khemani, S. P. Lim, D. N. Sheng, D. A. Huse,arxiv:1607.05756], it seems thus useful to introduce the notion of a finite-size-pseudo-critical point for each individual eigenstate and to study its eigenstate-to-eigenstate fluctuations governed by the correlation length exponent $\nu$.  The scaling properties of critical eigenstates are also expected to appear much more clearly if one considers each eigenstate at its finite-size-pseudo-critical point, where it is 'truly critical', while standard averages over eigenstates and samples in the critical region actually see a mixture of states that are effectively either localized or delocalized.

\end{abstract}

\maketitle

\section{ Introduction} 

In the field of phase transitions for pure models, the finite-size-scaling theory is essential to extract from numerical studies
on finite sizes $L$ the location $\theta_c$ of the critical point in the thermodynamic limit $L \to +\infty$ and the critical exponents of various observables \cite{cardyFSS} : the main idea is that the important variable is
the ratio between the size $L$ of the finite system and
the correlation length that diverges $\xi(\theta) \sim \vert  \theta-\theta_c \vert^{-\nu}$ as the control parameter $\theta$ approaches the critical value $\theta_c$.
One possible rephrasing is that the finite sample of size $L$ becomes critical when the correlation length $\xi(\theta)$ reaches $L$
and thus its finite-size-pseudo-critical point $\theta_c(L)$ defined by $\xi(\theta_c(L)) =L$ converges towards its thermodynamic limit $\theta_c$ as
\begin{eqnarray}
\theta_c(L) - \theta_c \opsimeq   L^{-\frac{1}{\nu}}
\label{pure}
\end{eqnarray}
In the presence of quenched disorder, one should take into account the additional property that a given size $L$ does not correspond to a single sample anymore, but to the whole set of all possible disordered samples of size $L$. The most widespread way to deal with this problem is to compute the disorder-averaged values of observables and to analyze them with the standard finite-size-scaling theory of pure models.
However in the critical region, sample-to-sample fluctuations are expected to be large \cite{domany95,AH,paz1,domany,AHW,paz2,chamati}. For instance at some given size $L$, some fraction of the samples seem in one phase while the complementary fraction seem in the other phase. To better take into account these sample-to-sample fluctuations,
 it has been proposed to associate to each disordered sample $(\omega)$ of size $L$ its own finite-size pseudo-critical point $\theta_c(\omega,L)$ \cite{domany95,paz1,domany,paz2,us_ihes,us_ihp} : the goal is to try to separate
the finite-size-scaling $ (\theta-\theta_c(\omega,L) )L^{\frac{1}{\nu}}$ within a given sample from the sample-to-sample fluctuations of pseudo-critical point $\theta_c(\omega,L)$.
One possibility to define $\theta_c(\omega,L)$ is to use the location of the maximum of some observable that is expected to diverge in the thermodynamic limit, but many other definitions can actually be used depending on the model and on the observables that are numerically measured. The important point is that 
the scaling properties of the distribution of this finite-size pseudo-critical point $\theta_c(\omega,L)$
should not depend on the precise details of the definition.
For conventional random critical points described by a single correlation exponent $\nu$,  one expects that Eq. \ref{pure} becomes
\begin{eqnarray}
\theta_c(\omega,L) - \theta_c \opsimeq  x_{\omega} L^{-\frac{1}{\nu}}
\label{random}
\end{eqnarray}
where $x_{\omega} $ is an $O(1)$ sample-dependent random variable.
The probability distribution of the pseudo-critical points $ \theta_c(\omega,L) $ 
has been studied in many disordered classical models involving either spins \cite{paz1,domany,paz2,igloipottsq,castel,billoire}, 
elastic lines in random media \cite{bolech}, or disordered polymers 
\cite{PS2005,interface2005,barbara}, as well as in dynamical epidemic models in random media \cite{contact}.
It has also been very much used for
quantum phase transitions concerning the ground state of disordered quantum spin models
in various dimensions \cite{us_tfic,kovacsstrip,kovacs2d,kovacs3d,kovacsreview,kovacsLR,kovacsLR3d}.

The aim of the present paper is to discuss this notion of finite-size pseudo-critical points for the case of the
Many-Body-Localization (MBL)  Transition where one is interested into the properties of the excited eigenstates of interacting disordered quantum models (see the recent reviews \cite{revue_huse,revue_altman,revue_vasseur,revue_imbrie,review_mblergo,review_rare} and references therein). In the Ergodic phase where the Eigenstate Thermalization Hypothesis (E.T.H.) \cite{deutsch,srednicki,nature,mite,rigol} holds, the eigenstates display the volume-law entanglement with a prefactor given by the thermal entropy.
In the Many-Body-Localized (MBL) phase, the eigenstates display an area-law entanglement \cite{bauer,grover,kjall,alet,alet_dyn,luitz_tail,badarson_signa} with a power-law entanglement spectrum \cite{serbyn_powerlawent} and there exists an extensive number 
of emergent localized conserved operators \cite{emergent_swingle,emergent_serbyn,emergent_huse,emergent_ent,
imbrie,serbyn_quench,emergent_vidal,emergent_ros,emergent_rademaker}. 
The critical point between these two phases remains not well understood, because it is definitely unusual from various points of view.
In particular, very strong eigenstate-to-eigenstate fluctuations 
 even within a given disordered sample at a given energy density have been reported recently \cite{luitz_bimodal,huse_eigendep}.
This is a very surprising result with respect to the 'self-averaging' feature that was always taken for granted up to now.
The physical interpretation given in \cite{huse_eigendep}
is that the transition is driven by an eigenstate-dependent sparse resonant backbone.
 In this paper, we analyze various models via strong disorder expansion 
in order to show that it makes sense to associate a finite-size-pseudo-critical point 
$\theta_c(n,\omega,L)$ to each eigenstate $n$ of a given disordered sample $\omega$ (and not to each disordered sample $\omega$ as discussed above for thermal classical transitions or for quantum phase transitions concerning only the ground state).

The paper is organized as follows.
The idea of a finite-size-pseudo-critical point for each eigenstate 
is first discussed for the simpler case of the Anderson Localization transition concerning a single particle,
either with long-ranged hopping in section \ref{sec_anderson}, or with nearest-neighbor hopping on the Cayley tree in section \ref{sec_cayley}.
In section \ref{sec_toy}, we analyze in detail the MBL quantum spin chain toy model of Ref \cite{c_mblper}.
Finally in section \ref{sec_heisen}, we consider the standard model of MBL, namely the nearest-neighbor Heisenberg chain in random fields.
Our conclusions are summarized in section \ref{sec_conclusion}.

\section{ Anderson Localization transition with long-ranged hoppings  } 

\label{sec_anderson}

\subsection{ Reminder on the one-dimensional power-law hopping model }

At Anderson localization transitions, 
the critical eigenstates can be more or less multifractal 
(see the review \cite{mirlinrevue} and references therein).
For the short-ranged tight-binding model in dimension $d$, 
there is a continuous interpolation 
between the 'weak multifractality' regime in $d=2+\epsilon$ and the 'strong multifractality' in high dimension $d$.
For the one-dimensional tight-binding model 
\begin{eqnarray}
H && = \sum_{n} \epsilon_n \vert n > <  n \vert + \sum_{n \ne m} 
V_{nm}  \vert n > <  m \vert 
\label{anderson}
\end{eqnarray}
with random on-site energies $\epsilon_n$ and
 power-law hopping with respect to the distance
\begin{eqnarray}
V_{nm}  =  V \frac{  v_{nm}}{\vert n-m\vert^a} 
\label{hopping}
\end{eqnarray}
where the $v_{nm}$ of order $O(1)$ can be either fixed or random, 
and where the prefactor $V$ is the global amplitude.
The control parameter of the transition
 is the exponent $a$ governing the decay with the distance in Eq \ref{hopping}, and 
the critical point in the thermodynamic limit is exactly known to be
\begin{eqnarray}
a_c=1
\label{acriti}
\end{eqnarray}
As a function of the amplitude $V$, 
 the universality class of the critical point varies continuously 
 from the 'weak multifractality' regime and the nearly Wigner Dyson statistics for large $V \to +\infty$ 
to the 'strong multifractality' regime and the nearly Poisson statistics for small $V \to 0$ \cite{mirlinrevue}.
This 'strong multifractality'
regime $ V \to 0$ has been analyzed via the 
Levitov renormalization method that takes into account the
resonances occuring at various scales \cite{levitov1,levitov2,levitov3,levitov4,mirlin_evers,fyodorov,fyodorovrigorous}
or other methods \cite{oleg1,oleg2,oleg3,oleg4,olivier_per,olivier_strong,olivier_conjecture}, including 
 first-order perturbation theory of quantum mechanics \cite{us_strongmultif} that we use in the following.

\subsection{ Finite-size-pseudo-critical associated to each eigenstate }

Let us now consider a given sample containing $L$ sites on a ring, where the $L$ random on-site energies 
$\epsilon_n$ drawn with some probability distribution $P(\epsilon_n)$ are given.
Following \cite{us_strongmultif}, 
we consider the perturbation theory with respect to the hopping amplitude $V \to 0$.
At order zero, the eigenstates $\vert \phi_{n}^{(0)} > $ are completely localized on a single site
\begin{eqnarray}
\vert \phi_{n}^{(0)} > && = \vert n > 
\label{phi0}
\end{eqnarray}
 and the corresponding eigenvalues $E_n$
are given by the on-site energies
\begin{eqnarray}
E^{(0)}_{n} && = \epsilon_{n}
\label{ei0}
\end{eqnarray}
At first order in the hopping amplitude $V \to 0$, 
the eigenvalues are unchanged
\begin{eqnarray}
E_n^{(0+1)}  = \epsilon_{n}
\label{ener1}
\end{eqnarray}
while the eigenstates become
\begin{eqnarray}
\vert \phi_n^{(0+1)} > && 
  = \vert n> + \sum_{ m \ne n} R_{nm} \vert m > 
\label{phi1}
\end{eqnarray}
in terms of the hybridization ratios
\begin{eqnarray}
R_{nm} = \frac{V_{nm} }{ \epsilon_n- \epsilon_m} =   \frac{V v_{nm}}{\vert n-m\vert^a  (\epsilon_n- \epsilon_m)}
\label{rij}
\end{eqnarray}
We refer to \cite{us_strongmultif} for the explicit computations concerning the Inverse Participation Ratios
governed by the strong multifractality spectrum.
Here as explained in the Introduction, our goal is to define a finite-size-pseudo-critical point $a_c(n,\omega,L)$
for each eigenstate $\vert \phi_n > $ of the disordered sample $\omega$ of length $L$.

To detect the transition, it is actually sufficient to consider
the hybridization ratio in Eq. \ref{phi1}
corresponding to the index $m=m(n)$ whose energy $\epsilon_{m(n)} $  is the closest to the energy $\epsilon_n$,
i.e. the most 'dangerous' resonance
\begin{eqnarray}
\vert \epsilon_n- \epsilon_{m(n)} \vert = 
\opmin_{m \ne n} \vert \epsilon_n- \epsilon_{m} \vert
\label{ediff}
\end{eqnarray}
This energy difference can be rewritten as
\begin{eqnarray}
\vert \epsilon_n- \epsilon_{m(n)} \vert = \Delta_L(\epsilon_n)  s_n
\label{si}
\end{eqnarray}
where $s_n$ is a random variable of order $O(1)$, and where the characteristic scale of the level spacing around the energy $\epsilon_n$ reads
\begin{eqnarray}
\Delta_L(\epsilon_n) = \frac{1}{L P(\epsilon_n) }
\label{deltai}
\end{eqnarray}
Since the site $m(n)$ has been chosen on the purely energetic criterion of Eq. \ref{si} concerning $\epsilon_{m(n)} $,
the position $m(n)$ is uniformly distributed among the $(L-1)$  sites different from $n$, so that the distance 
$l_n = \vert n-m(n)\vert $ on the ring can be replaced by 
\begin{eqnarray}
l_n = L u_n
\label{uniform}
\end{eqnarray}
where $u_n$  is a uniform variable on $[\frac{1}{L},\frac{1}{2}]$.
Putting everything together, 
the hybridization ratio $\vert R_{n, m(n)} \vert $ of $n$ with the closest-energy resonance of Eq. \ref{ediff}
reads
\begin{eqnarray}
 \vert R_{n, m(n)} \vert =   \frac{V \vert v_{n,m(n)} \vert}{\vert n-m(n) \vert^{a} 
\vert \epsilon_n- \epsilon_{m(n)} \vert}
=   \frac{V \vert v_{n,m(n)} \vert  P(\epsilon_n)}{u_n^{a}  s_n } L^{1-a} 
\label{rinext}
\end{eqnarray}
As a consequence of the dependence as $ L^{1-a} $ with respect to the size $L$,
this hybridization ratio diverges $R \to +\infty$ in $L$ in the delocalized phase $a<a_c=1$,
 vanishes $R \to 0$ in $L$ in the localized phase $a>a_c=1$, and remains distributed at criticality $a_c=1$.
So for finite $L$, we can choose to define the finite-size-pseudo-critical point $a_c(n,\omega,L)$ for the eigenstate $\vert \phi_n > $
as the value of the control parameter $a$ where 
the hybridization ratio of Eq. \ref{rinext}
takes the value unity
\begin{eqnarray}
1= \vert R_{n, m(n)} \vert =   \frac{V \vert v_{n,m(n)} \vert  P(\epsilon_n) L }{(u_nL)^{a_c(n,\omega,L)}  s_n } 
\label{rinext1}
\end{eqnarray}
leading to
\begin{eqnarray}
a_c(n,\omega,L)   
 = a_c + \frac{ x_n  }{\ln L }   + o \left(  \frac{1}{ \ln L}\right)
\label{acilexp}
\end{eqnarray}
in terms of the infinite-size transition location $a_c=1$ of Eq. \ref{acriti}
and of the $O(1)$ random variable associated to the eigenstate $\vert \phi_n >$
\begin{eqnarray}
x_n && \equiv \ln  \left( \frac{V \vert v_{n,m(n)} \vert P(\epsilon_n)}{  s_n u_n } \right)
\label{vianderson}
\end{eqnarray}
Eq \ref{acilexp} follows the general form of Eq. \ref{random} with an infinite correlation exponent $\nu=\infty$
corresponding to the exponential divergence of the correlation length
 of this model \cite{mirlin_power,power_wegnerflow}
\begin{eqnarray}
\ln \xi(a) \propto \frac{1}{\vert a-a_c \vert}
\label{xia}
\end{eqnarray}

In summary, on the example of the one-dimensional power-law Anderson Localization model in the strong multifractality regime, 
we have shown that it makes sense to associate to each eigenstate its own finite-size-pseudo-critical point $a_c(n,\omega,L)$
and that its probability distribution is described by Eq. \ref{acilexp}, so that it allows to identify the 
thermodynamic critical point $a_c=1$ and the correlation length exponent $\nu$.

\section{ Anderson nearest-neighbor model on the Cayley tree  }

\label{sec_cayley}

In this section, we consider the finite Cayley tree of branching ratio $K$ and containing $L$ generations besides the central root $O$ :
there are $(K+1) K^{n-1}$ sites on the generation $n$, so that the total number of sites is
\begin{eqnarray}
{\cal N}_L = 1+\sum_{n=1}^L (K+1) K^{n-1} = 1+(K+1) \frac{K^L-1}{K-1} 
\label{nlcayley}
\end{eqnarray}

For the Anderson Localization model with hopping $V$ between nearest-neighbors
and with random on-site energies $\epsilon_i$ drawn with the flat distribution of width $W$
\begin{eqnarray}
p(\epsilon) = \frac{ \theta( -W \leq \epsilon \leq W )}{2 W}
\label{flat}
\end{eqnarray}
 the critical point $V_c$ is known to be in the region of strong disorder 
when the branching ratio $K$ is large
\begin{eqnarray}
\frac{V_c}{W} \ \  \oppropto_{K \gg 1} \frac{1}{K \ln K} \ll 1
\label{zratio}
\end{eqnarray}
so that it makes sense to consider the perturbation theory in $V$.
However, in contrast to the previous sections with long-ranged hoppings,
 the first-order perturbation theory in $V$ is of course not sufficient
to reach all configurations of the Hilbert space,
 so that one needs to use the so-called
Forward Approximation \cite{alt_levitov,luca,forward}
as we now recall.

\subsection{Forward Approximation for an eigenstate   } 

At order zero in the hopping $V=0$,  
the eigenstates are completely localized
 on the sites, and the corresponding eigenvalues are simply the
 random on-site energies $\epsilon_i$
\begin{eqnarray}
\vert \phi_i^{(0)} > && = \vert i > \nonumber \\
E_i^{(0)} && = \epsilon_{i}
\label{cphi0}
\end{eqnarray}
 In the Forward Approximation \cite{alt_levitov,luca,forward}, one focuses on the eigenstate that is localized on the root 
$\vert \phi_0^{(0)} >  = \vert 0 >  $  at order zero,
and one writes the amplitudes at lowest order in perturbation theory with respect
to the hopping $V$ for all other sites of the tree  :
the $(K+1)$ sites $i_1=1,2,..,K+1$ of the first generation have amplitudes of order $V $
\begin{eqnarray}
 \phi_0^{(1)} (i_1) && = \frac{V}{\epsilon_0-\epsilon_{i_1}}
\label{phi01}
\end{eqnarray}
the $(K+1)K$ sites of the second generation have amplitudes of order $V^2$
\begin{eqnarray}
 \phi_0^{(2)} (i_1,i_2) && = \frac{V^2}{(\epsilon_0-\epsilon_{i_1})(\epsilon_0-\epsilon_{i_1,i_2})}
\label{phi02}
\end{eqnarray}
and so on up to the $(K+1) K^{L-1}$ sites of the last generation $L$ that have amplitudes of order $V^L $
\begin{eqnarray}
 \phi_0^{(L)} (i_1,i_2,..,i_L) && = \frac{V^L}{(\epsilon_0-\epsilon_{i_1})(\epsilon_0-\epsilon_{i_1,i_2}) ... (\epsilon_0-\epsilon_{i_1,i_2... i_L} ) }
\label{phi0l}
\end{eqnarray}
that involve all the on-site energies along the single path leading to the root.

\subsection{Finite-size pseudo-critical point for an eigenstate   }

For this eigenstate $\phi_{n=0}$ defined 
on the finite Cayley tree with $L$ generations with a realization $\omega$
of the ${\cal N}_L$ on-sites energies, 
one can define the pseudo-critical point $V_c(0,\omega,L)$ 
as the hopping $V$ where the maximum over the $(K+1)K^{L-1}\simeq K^L $
 amplitudes $\vert \phi_0^{(L)} (i_1,i_2,..,i_L)  \vert $ of the
 last generation $L$ reaches the value unity
\begin{eqnarray}
1= \left[ \frac{V_c(0,\omega,L)}{W} \right]^L   \opmax_{i_1,..i_L}   \left(  \frac{W^L}{\vert (\epsilon_0-\epsilon_{i_1})(\epsilon_0-\epsilon_{i_1,i_2}) ... (\epsilon_0-\epsilon_{i_1,i_2... i_L} )  \vert } \right)
\label{maxtree}
\end{eqnarray}

On the tree with large branching ratio $K \gg 1$, it turns out that
correlations in this Directed Polymer model are negligible \cite{alt_levitov,luca,forward},
so that Eq \ref{maxtree} can be replaced by the simpler problem
\begin{eqnarray}
1= \left[ \frac{V_c(0,\omega,L)}{W} \right]^L   \opmax_{1 \leq p \leq K^L}   \left( A_L(p) \right)
\label{maxtreea}
\end{eqnarray}
involving the maximum over $K^L$ independent variables $A_L(p)$ 
corresponding to the statistics of a path of length $L$ 
\begin{eqnarray}
A_L = \prod_{j=1}^L  \frac{W}{ \vert \epsilon_0- \epsilon_j \vert }
\label{amplial}
\end{eqnarray}

To simplify the notations from now on, we will consider that the on-site energy at the root is exactly at the center of the band
\begin{eqnarray}
\epsilon_0=0
\label{centerband}
\end{eqnarray}
Then the rewriting of Eq. \ref{amplial} in logarithmic variables
corresponds to a sum
\begin{eqnarray}
U_L \equiv \ln A_L = \sum_{j=1}^L  u_j
\label{loga}
\end{eqnarray}
of independent variables 
\begin{eqnarray}
u_j \equiv  \ln \left( \frac{W}{ \vert \epsilon_j \vert } \right)
\label{uj}
\end{eqnarray}
drawn with the exponential distribution (using Eq. \ref{flat})
\begin{eqnarray}
 P_1(u) && = \int_{-W}^{+W} \frac{d\epsilon }{2W}
\delta \left( u-  \ln \left( \frac{W}{ \vert \epsilon \vert } \right) \right)
 = e^{-u} \theta(u \geq 0)
\label{expx}
\end{eqnarray}
So the sum $U_L$ of $L$ such variables is distributed
 with the convolution of $L$ exponential distribution
\begin{eqnarray}
 P_L(U_L) && 
 = \frac{(U_L)^{L-1}}{(L-1)!} e^{-U_L} \theta(U_L \geq 0)
\label{expxconvol}
\end{eqnarray}
The average value is $\overline{U_L}=L$, 
but here we need to analyze the large deviation properties,
i.e. the exponentially small probability in $L$
 to have to have an anomalously large $y=\frac{U_L}{L}$
\begin{eqnarray}
 {\cal P}_L \left(y=\frac{U_L}{L} \right) &&  \propto e^{- L  I(y)}
\label{expxconvolrate}
\end{eqnarray}
where $I(y)$ is called the rate function (see the review on large deviations \cite{touchette}).
Using the Stirling formula
\begin{eqnarray}
(L-1)!   \opsimeq_{ L \to +\infty} \sqrt{2 \pi (L-1)} \left( \frac{L-1}{e} \right)^{L-1} 
\label{stirling}
\end{eqnarray}
one obtains from Eq. \ref{expxconvol}
\begin{eqnarray}
 {\cal P}_L \left(y=\frac{U_L}{L} \right) && 
 \propto \frac{(Ly)^{L-1}}{  \left( \frac{L-1}{e} \right)^{L-1}  } e^{- Ly }   
\label{expxconvollargedev}
\end{eqnarray}
so that the rate function of Eq. \ref{expxconvol} reads
\begin{eqnarray}
I(y)=y-1 -\ln y
\label{irate}
\end{eqnarray}

The probability distribution $Q_{max}(y_{max}) $
of the maximum $y_{max}$ among $K^L$ such variables $y$
can be obtained from the cumulative distribution
\begin{eqnarray}
\int^y dy_{max} Q_{max}(y_{max}) && = \left[ 1- \int_{y}^{+\infty} dy'  {\cal P}_L(y' ) \right]^{K^L}
\simeq e^{- K^L \int_{y}^{+\infty} dy'  {\cal P}_L(y' ) } \simeq e^{- K^L \int_{y}^{+\infty} dy'  e^{-L I(y') } }
\nonumber \\
&& \simeq  e^{-   e^{ L( \ln K - I(y) ) } }
\label{cumulqmax}
\end{eqnarray}
It is thus useful to introduce the value $y^*$ satisfying
\begin{eqnarray}
 \ln K = I(y^*) = y^*-1 -\ln y^*
\label{ystar}
\end{eqnarray}
and 
\begin{eqnarray}
b_L \equiv \frac{1}{L I'(y^*)} = \frac{1}{L \left[ 1-\frac{1}{y^*} \right] } 
\label{bldef}
\end{eqnarray}
Then the change of variables $y=y^*+b_L x$ in Eq. \ref{cumulqmax}
yields the convergence towards the Gumbel distribution for the $O(1)$ variable $x$
\begin{eqnarray}
\int^{y^*+b_L \xi} dy_{max} Q_{max}(y_{max}) 
&& \simeq  e^{-   e^{ L( \ln K - I(y^*)-b_L x I'(y^* )+O(b_L^2) ) } } \simeq e^{-   e^{ -  x  } } 
\label{cumulqmaxxi}
\end{eqnarray}

The pseudo-critical-point of Eq. \ref{maxtreea} becomes
\begin{eqnarray}
 \frac{V_c(0,\omega,L)}{W} 
&& = \left[  \opmax_{1 \leq p \leq K^L}   \left( A_L(p) \right)  \right]^{-\frac{1}{L}}
 = \left[  e^{L y_{max}}  \right]^{-\frac{1}{L}} \simeq e^{-y_{max}}  = e^{-y^*-b_L x}
\nonumber \\
&& =  e^{ - y^*- \left( \frac{ y^* }{ L (y^*-1) } \right) x } 
\opsimeq_{L \to +\infty}  e^{ - y^*} - \left( \frac{ y^* e^{ - y^*}}{ (y^*-1) } \right)\frac{ x  }{L}
\label{zctreedistri}
\end{eqnarray}
where $x$ is an $O(1)$ random variable drawn with the Gumbel distribution (Eq. \ref{cumulqmaxxi}). The convergence
 in $\frac{1}{L}$ corresponds to the correlation length exponent 
\begin{eqnarray}
\nu=1
\label{nutreethermo}
\end{eqnarray}
in agreement with the exact result for the correlation length exponent \cite{kunz,mirlin_fyodorov,us_andersontreeTW}.

In the thermodynamical limit $L \to +\infty$, the location of the transition 
is given by
\begin{eqnarray}
 \frac{V_c(L \to +\infty) }{W}   =  e^{ - y^* } 
\label{zctreethermo}
\end{eqnarray}
where $y^*$ is the solution of Eq. \ref{ystar}, that reads at leading order
 for large $K$ 
\begin{eqnarray}
y^*=  \ln K +\ln (e)+\ln y^*   = \ln ( Ke ) +\ln (\ln ( Ke )+\ln y^* ) 
 \simeq \ln ( Ke \ln ( Ke ) )
\label{ystarsolu}
\end{eqnarray}
leading to \cite{alt_levitov,luca,forward}
\begin{eqnarray}
 \frac{V_c (L \to +\infty) }{W}   =  e^{ - y^* }  \simeq \frac{1}{ Ke \ln ( Ke )}
\label{zctreethermosolu}
\end{eqnarray}

In summary, for the nearest-neighbor Anderson model on the Cayley tree,
 the forward approximation for the eigenstates allows to define a pseudo-critical point $V_c(0,\omega,L )$
that allows to identify the thermodynamic critical point $V_c (L \to +\infty)$ and the correlation length exponent $\nu=1$.

\section{ MBL quantum spin chain toy model of Ref \cite{c_mblper}    } 

\label{sec_toy}

In this section, we consider the MBL quantum spin chain toy model that has been introduced in \cite{c_mblper}
in direct analogy with the power-law hopping Anderson model described in section \ref{sec_anderson}.

\subsection{ Unperturbed Hamiltonian $H_0$ with completely localized eigenstates }

The unperturbed Hamiltonian 
\begin{eqnarray}
H_0 && = - \sum_{j=1}^{L} h_j \sigma_j^z
\label{H0}
\end{eqnarray}
contains only $N$ random fields $h_j$.
The $2^{L}$ corresponding eigenstates are simply the tensor products
\begin{eqnarray}
\vert \psi^{(0)}_{S_1,..,S_{L}} > && \equiv \vert S_1 > \otimes \vert S_2 > ... \otimes \vert S_{L} >
\label{psizero}
\end{eqnarray}
with the random energies
\begin{eqnarray}
E^{(0)}_{S_1,..,S_{L}} =- \sum_{j=1}^{L} h_j S_j
\label{e0}
\end{eqnarray}
For instance, the ground state corresponds to the choice $S_j={\rm sgn} (h_j)$
and has the extensive energy
\begin{eqnarray}
E^{(0)}_{GS} = - \sum_{j=1}^{L} \vert h_j \vert 
\label{egs}
\end{eqnarray}

Here we wish to consider a given sample $\omega$ where the $L$ random fields $h_j$ are fixed,
so that the density of states $\rho_{\omega}(E) $ in this sample $\omega$
\begin{eqnarray}
\rho_{\omega}(E) \equiv \frac{1}{2^L} \sum_{ E^{(0)}_{S_1,..,S_{L}}} \delta  \left(  E- E^{(0)}_{S_1,..,S_{L}} \right)
= \frac{1}{2^L} \sum_{ S_1=\pm 1, .., S_L=\pm 1}  \delta  \left(  E +\sum_{j=1}^{L} h_j S_j\right)
\label{rhoh}
\end{eqnarray}
has for Fourier transform
\begin{eqnarray}
{\hat \rho}_{\omega}(k) \equiv \int dE \rho_{\omega}(E) e^{i k E} 
= \prod_{j=1}^L \cos( k h_j ) 
\label{rhohfourier}
\end{eqnarray}

For large $L$,  one recovers of course the Central Limit Gaussian form around the origin
\begin{eqnarray}
{\hat \rho}_{\omega}(k) && \opsimeq_{L \to +\infty} e^{- \frac{k^2}{2} \sigma_{\omega}^2 }
\nonumber \\
\rho_{\omega}(E)  && \opsimeq_{L \to +\infty}  \frac{1}{\sqrt { 2  \pi L \sigma_{\omega}^2 }} e^{- \frac{ E^2}{2  L \sigma_{\omega}^2} }
\label{gaussrho}
\end{eqnarray}
involving the sample-parameter
\begin{eqnarray}
\sigma_{\omega}^2 \equiv \frac{1}{L} \sum_{j=1}^L h_j^2 
\label{varmiddle}
\end{eqnarray}
that represents the effective variance seen by the given sample $\omega$.
Since it is the rescaled sum of $L$ independent random variables, its distribution over the samples is also governed by the
Central Limit Theorem : it converges towards the variance $\overline{h_j^2}$ in the thermodynamic limit $L \to +\infty$ for all samples, 
but otherwise displays $\frac{1}{\sqrt L}$ fluctuations for finite size $L$
\begin{eqnarray}
\sigma_{\omega}^2  \opsimeq_{L \to +\infty} \overline{h_j^2} + \frac{y_{\omega}}{\sqrt{L}}
\label{varmiddleclt}
\end{eqnarray}
where $y_{\omega}$ is a Gaussian random variable associated to the sample $\omega$.
Since there are $2^L$ levels, the level spacing near zero energy can be obtained from Eq. \ref{gaussrho} as
\begin{eqnarray}
\Delta_{\omega,L}(E=0)  \simeq \frac{1}{2^L \rho_{\omega}(E=0)} = \sqrt { 2  \pi L \sigma_{\omega}^2 }   \ \  2^{-L}
\label{levelspacing}
\end{eqnarray}

Beside the center of the spectrum just discussed,
it is also interesting to consider the eigenstates with a given energy density
\begin{eqnarray}
e = \frac{E}{L} 
\label{edensity}
\end{eqnarray}
in order to analyze the presence of some mobility edge in energy.
For instance in the small deviation region where Eq. \ref{gaussrho} can still be used, one obtains
\begin{eqnarray}
\rho_{\omega}(E=L e) \simeq  \frac{1}{\sqrt { 2  \pi L \sigma_{\omega}^2 }} e^{- L \frac{ e^2}{2  \sigma_{\omega}^2} }
\label{rhohe}
\end{eqnarray}
and the corresponding level spacing
\begin{eqnarray}
\Delta_{\omega,L}(E=Le)  \simeq \frac{1}{2^L \rho_{\omega}(E=Le)} = \sqrt { 2  \pi L \sigma_{\omega}^2 }    e^{-L \left( \ln 2 -  \frac{ e^2}{2  \sigma_{\omega}^2}  \right) }
\label{levelspacinge}
\end{eqnarray}
while for larger $\vert e \vert$ one should use the large-deviation theory (see for instance the review \cite{touchette} and references therein)
to obtain from Eq. \ref{rhohfourier}
the entropy $S_{\omega}(e)$ 
that governs the leading exponentially small term of the level spacing
\begin{eqnarray}
\Delta_{\omega,L}(E=Le)  \oppropto   e^{-L S_{\omega}(e) }
\label{largedev}
\end{eqnarray}
The entropy behaves quadratically near the origin (Eq. \ref{levelspacinge})
\begin{eqnarray}
S_{\omega}(e) \opsimeq_{e \to 0} \ln 2 -  \frac{ e^2}{2  \sigma_{\omega}^2} 
\label{entropy}
\end{eqnarray}
and vanishes beyond the energy density of the ground state (Eq. \ref{egs})
\begin{eqnarray}
 e^{GS}_{\omega} = - \frac{1}{L}  \sum_{j=1}^{L} \vert h_j \vert 
\label{segs}
\end{eqnarray}
that also displays $\frac{1}{\sqrt L}$ sample-to-sample fluctuations around 
its thermodynamic limit $(- \overline{ \vert h_j \vert})$.

\subsection{ Perturbation $H_1$  } 

In analogy with Eq. \ref{anderson},
the small perturbation is chosen to produce a direct coupling between all pairs of states the Hilbert space
\cite{c_mblper}
\begin{eqnarray}
H_1 && = - \sum_{k=1}^{L}  \sum_{1 \leq i_1< i_2..<i_k \leq {L}} 
J_{i_1,..,i_k}  \sigma^x_{i_1} \sigma^x_{i_2} ... \sigma^x_{i_k}
\label{Vperturbation}
\end{eqnarray}
and the couplings $ J_{i_1,..,i_k} $ are chosen in analogy with Eq. \ref{hopping}
\begin{eqnarray}
J_{i_1,..,i_k} = V \  \frac{2^{-b \vert i_k - i_1 \vert} v_{i_1,..,i_k}}{ \vert i_k - i_1 \vert^a} 
\label{gaussJ}
\end{eqnarray}
where the $v_{i_1,..,i_k}$ are $O(1)$ random variables,
and where the prefactor $V$ is the global amplitude.
The decay with respect to the spatial range $r=i_k-i_1$ 
contains the leading exponential decay governed by the control parameter $b$
and possibly some power-law prefactor governed by the parameter $a$.

At first order in the perturbation $H_1$, the eigenvalues of Eq. \ref{e0} are unchanged 
 \begin{eqnarray}
E^{(0+1)}_{S_1,..,S_{L}} = E^{(0)}_{S_1,..,S_{L}} 
\label{e1}
\end{eqnarray}
while the eigenstates read
 \begin{eqnarray}
  \vert \psi^{(0+1)}_{S_1,..,S_L} >  =  \vert S_1,..,S_L >
+ \sum_{ S_1'=\pm 1, ...S_L' =\pm 1 } R \left(S_1..S_L \vert S_1'...S_L' \right) \vert S_1',..,S_L' >
 \label{eigenzeroper}
  \end{eqnarray}
in terms of the hybridization ratios
\begin{eqnarray}
R\left(S_1..S_L \vert S_1'...S_L' \right) \equiv \frac{ < S_1',..,S_L' \vert H_1 \vert S_1,..,S_L > }
{ E^{(0)}_{S_1,..,S_L} -  E^{(0)}_{S_1',..,S_L'}}
\label{rsisip}
\end{eqnarray}
We refer to \cite{c_mblper} for the explicit computations concerning 
the multifractality of the entanglement spectrum in the localized phase and at criticality.
Here as explained in the Introduction, our goal is to define a finite-size-pseudo-critical point $b_c( n,\omega,L  )$
for each eigenstate $\vert n> \equiv\vert \psi^{(0+1)}_{S_1,..,S_L} >   $ of the sample $\omega$ of length $L$.

\subsection{ Finite-size-pseudo-critical point for each eigenstate }

As in the Anderson case, it is sufficient to consider the hybridization ratio of Eq. \ref{eigenzeroper}
between the state $ n\equiv (S_1,...,S_L)$
and the configuration $m(n) \equiv (S_1',...,S_L')$ whose energy $E^{(0)}_{S_1',..,S_L'}$
is the closest to $ E^{(0)}_{S_1,..,S_{L}} $ : then the energy difference reads
\begin{eqnarray}
\opmin_{S_1'...,S_L'} \vert E^{(0)}_{S_1,..,S_L} -  E^{(0)}_{S_1',..,S_L'}\vert =s_n  \Delta_{\omega,L}(E_n)  
\label{sim}
\end{eqnarray}
in terms of the the level spacing  $\Delta_{\omega,L}(E_n) $ around the energy $E_n \equiv E^{(0)}_{S_1,..,S_{L}}$ 
 and of the $O(1)$ random variable $s_n$.

Since the configuration $(S_1',...,S_L')$ has been chosen on the purely energetic criterion of Eq. \ref{sim},
it is expected to involve typically $k \simeq \frac{L}{2}$ spin flips,
and the corresponding spatial range between the locations $i_1$ and $i_k$ of the first spin and the last spin flips
is given by the system size $L$ (see more details on the properties of resonances in \cite{c_mblper})
\begin{eqnarray}
r= \vert i_k - i_1 \vert  \simeq L - o(L)
\label{rmaxi}
\end{eqnarray}
Indeed, the Hilbert space of configurations limited to a spatial range $r=\alpha L$
grows as $2^r$, with a level spacing decaying as $2^{-r}=2^{-\alpha L}$,
so that it corresponds to the level spacing $2^{-L}$ of the total system only for $\alpha=1$.

So that the off-diagonal matrix element of the numerator of Eq. \ref{rsisip}
is given by some coupling (Eq.\ref{gaussJ}) of maximal range 
\begin{eqnarray}
< S_1',..,S_L' \vert H_1 \vert S_1,..,S_L >= \frac{V}{ L^a  } 2^{-b L  } v_n
\label{gaussJres}
\end{eqnarray}
where $v_n \equiv v_{i_1,..,i_k}  $ is the corresponding random $O(1)$ variable (Eq. \ref{gaussJ}).

Putting everything together, the hybridization ratio of Eq. \ref{rsisip} with the closest energy level becomes
\begin{eqnarray}
R_{n,m(n)} = \frac{ V v_n   2^{-b L  }}
{ s_n L^a \Delta_{\omega,L}(E_n)  }  
\label{rdan}
\end{eqnarray}

To be more explicit, let us focus on the small-deviation region regime around the middle of the spectrum
where the level spacing around the energy $E_n=E^{(0)}_{S_1,..,S_{L}}=Le$ of the eigenstate $ \vert \psi^{(0+1)}_{S_1,..,S_L} >$ is described by Eq. \ref{levelspacinge}
\begin{eqnarray}
\Delta_{\omega,L}(E_n=Le)  =
\sqrt { 2  \pi L \sigma_{\omega}^2 }    e^{-L \left( \ln 2 -  \frac{ e^2}{2  \sigma_{\omega}^2}  \right) }
\label{levelspacinge0}
\end{eqnarray}
so that Eq. \ref{rdan} becomes
\begin{eqnarray}
R_{n,m(n)} = \frac{ V v_n   }
{ s_n L^{a+\frac{1}{2}}  \sqrt { 2  \pi \sigma_{\omega}^2 }  }  e^{L \left( (1-b) \ln 2 -  \frac{ e^2}{2  \sigma_{\omega}^2}  \right) }
\label{rdanbis}
\end{eqnarray}
In the thermodynamic limit $L \to +\infty$, the behavior is governed by the mobility edge
given in the quadratic small-deviation region by
\begin{eqnarray}
b_c(e,\omega) \opsimeq_{e \to 0} 1 - \frac{ e^2}{2  \sigma_{\omega}^2  \ln 2 }
\label{bce}
\end{eqnarray}
in agreement with the value at the middle of the spectrum already studied in \cite{c_mblper}
\begin{eqnarray}
b_c(e=0) = 1 
\label{bce0}
\end{eqnarray}
while in the large-deviation region described by Eq. \ref{largedev} involving the entropy $S_{\omega}(e)$ of eigenstates of a given energy density $e$, it becomes
\begin{eqnarray}
b_c(e,\omega) =  \frac{S_{{\omega}}(e)}{\ln 2}
\label{bcelarge}
\end{eqnarray}

The ratio of Eq. \ref{rdanbis} diverges exponentially in $L$ in the delocalized phase $b<b_c(e)$,
vanishes exponentially in $L$ in the localized phase $b>b_c(e)$, 
and behaves as a power-law $ L^{-a-\frac{1}{2}} $
with a random prefactor at the critical point $b=b_c(e) $.
So let us define the finite-size-pseudo-critical point $b_c(n,\omega,L)$ of the eigenstate $n$
as the value of the control parameter $b$ where the hybridization ratio of Eq. \ref{rdan}
takes the value $L^{-a-\frac{1}{2}}  $ (i.e. the critical dependence in $L$ with amplitude unity)
\begin{eqnarray}
\frac{1}{ L^{a+\frac{1}{2}} }=  \frac{ V v_n   }
{ s_n L^{a+\frac{1}{2}}  \sqrt { 2  \pi \sigma_{\omega}^2 }  }  2^{L  (b_c(e)-b_c( n,\omega,L) )  }
\label{rinext1a}
\end{eqnarray}
leading to
\begin{eqnarray}
b_c( n,\omega,L) 
= b_c(e,\omega)  + \frac{ x_{n,\omega} }{ L } 
\label{bcil}
\end{eqnarray}
with the $O(1)$ random variable
\begin{eqnarray}
x_{n,\omega} \equiv  \frac{    \ln \left( \frac{V \vert v_n \vert }{ s_n   \sqrt { 2  \pi  \sigma_{\omega}^2 }     } \right) }{ \ln 2} 
\label{x0mbl}
\end{eqnarray}
Eq \ref{bcil} is thus of the form of Eq. \ref{random} with the correlation length exponent
\begin{eqnarray}
\nu = 1
\label{nu1}
\end{eqnarray}
in agreement with the previous study at the middle of the spectrum  \cite{c_mblper}.

Both the location $b_c(e)$ of the critical point and the value $\nu=1$ 
can be understood from the interpretation of the transition
as the crossing of the exponential decay of the level spacing of Eq. \ref{levelspacing} 
and of the exponential decay of the couplings of Eq. \ref{gaussJ}.

\subsection{ Effect of sample-to-sample fluctuations }

Up to now, we have considered that the disordered sample $\omega$ with its $L$ random fields $h_j$
was fixed, so that the entropy $S_{\omega}(e) $ determining the mobility edge in Eq. \ref{bcelarge} was fixed,
and in particular its small-fluctuation region of Eq. \ref{bce} involving the effective variance $ \sigma_{\omega}^2$
defined by Eq. \ref{varmiddle}.
However this parameter will fluctuate from sample-to-sample according to
the Central Limit Theorem of Eq. \ref{varmiddleclt}
\begin{eqnarray}
\sigma_{\omega}^2  \opsimeq_{L \to +\infty} \overline{h_j^2} + \frac{y_{\omega}}{\sqrt{L}}
\label{varmiddlecltbis}
\end{eqnarray}
where $y_{\omega}$ is a random variable associated to the sample $ {\omega}$.
So here a very important difference arises between zero and non-zero energy density :

(i) for zero-energy-density $e=0$ at the middle of the spectrum, the sample-parameter $\sigma_{\omega}^2$
does not appear in the mobility edge (Eq \ref{bce}) but only in the random variable $x_n$ of Eq. \ref{x0mbl},
so that the fluctuations of Eq. \ref{varmiddlecltbis} only give terms of higher order in Eq. \ref{bcil}.
As a consequence, when the eigenstates of zero-energy-density $e=0$ coming from various samples $\omega$ 
are put together, their finite-size-pseudo-critical point follows (Eqs \ref{bcil} and \ref{x0mbl})
\begin{eqnarray}
b_c( n,e=0,L) = 1  + \frac{ x_{n} }{ L } 
\label{bcile0}
\end{eqnarray}
with the $O(1)$ random variable
\begin{eqnarray}
x_{n} \equiv  \frac{    \ln \left( \frac{V \vert v_n \vert }{ s_n   \sqrt { 2  \pi  (\overline{h_j^2}) }     } \right) }{ \ln 2} 
\label{x0mblmiddle}
\end{eqnarray}
i.e. Eq \ref{bcile0} involves the same exponent $\nu=1$ as in Eq. \ref{nu1}.

(ii) for non-zero-energy-density $e \ne 0$, the sample-parameter $\sigma_{\omega}^2$
does appear in the mobility edge (Eq \ref{bce}) 
so that the fluctuations of Eq. \ref{varmiddlecltbis} 
induce sample-to-sample fluctuations of order $1/\sqrt{L}$ in the mobility edge
\begin{eqnarray}
b_c(e) \simeq 1 - \frac{ e^2}{2   \ln 2  \left(  \overline{h_j^2} + \frac{y_{\omega}}{\sqrt{L}}  \right) }
\simeq  1 - \frac{ e^2}{2   \ln 2  \overline{h_j^2}   }
+ \left( \frac{ e^2 y_{\omega} }{2   \ln 2  ( \overline{h_j^2} )^2  }  \right)
 \frac{ 1}{  \sqrt{L}}  
\label{bceexp}
\end{eqnarray}
These disorder Central-Limit-Fluctuations of order $ L^{-\frac{d}{2}}$ (here in dimension $d=1$)
are of course very well known in the field of phase transitions in random models :
they appear in the Harris criterion $\nu_{pure} > 2/d$ \cite{harris} 
for the stability of a pure critical point with respect to weak disorder;
they also appear in the Chayes-Chayes-Fisher-Spencer general bound for random critical points
\begin{eqnarray}
\nu_{FS} \geq \frac{2}{d}
\label{nufs}
\end{eqnarray}
The link between the Harris criterion and the Chayes-Chayes-Fisher-Spencer general bound
can be understood as follows : Eq. \ref{nufs} essentially means
 that a random critical point should itself be stable with respect to the a small change of disorder realization.
The bound of Eq. \ref{nufs} has been rediscussed recently for the specific case of the MBL transition \cite{harrisMBL}.

However, as explained in \cite{chayes}, the general bound of Eq. \ref{nufs} can have two very different meanings :

(a)  In so-called `conventional' random critical points, 
there is a single correlation length exponent
 $\nu=\nu_{FS}$ and this single exponent should satisfy the bound of Eq. \ref{nufs}.

(b)  In `unconventional' random critical points,  two different correlation length exponents coexist :
then the typical correlation exponent $\nu_{typ}$ can be less than $2/d$,
 while the bound holds for the finite-size exponent $\nu_{FS} \geq 2/d$.
 The best known example is the quantum phase transition of the random transverse field Ising chain 
exactly solved by Daniel Fisher via Strong Disorder Renormalization \cite{danielrtfic}, where
 the typical correlation exponent $ \nu_{typ}=1$ is less than $2/d=2$,
 while the finite-size exponent $\nu_{FS}=2$ actually saturates the bond $\nu_{FS} = 2/d=2$.
 Another important example discussed in \cite{chayes,danielrtfic}
 is the case of a first order transition that remains first order in the presence of quenched disorder:
 this first order transition in dimension $d$ is associated to the typical exponent $\nu_{typ}=1/d$,
 which is less than $2/d$, whereas the finite-size exponent saturates the bound $\nu_{FS}=2/d$. 
 The interpretation given in Sec. VII A of Ref. \cite{danielrtfic}
 is that the exponent $\nu_{typ}=1/d$
is expected to describe the rounding of the transition in a 
typical sample, whereas $\nu_{FS}=2/d$ describes the rounding of the transition
of the distribution of samples.  
Other critical points with two different correlation length exponents are discussed in \cite{singh,paz1,fisher2nu,bolech,myers,PS2005,us_wettingtree}.
 
So for our present MBL model, our conclusion is that the finite-size exponent
\begin{eqnarray}
\nu_{FS} (e \ne 0)=2
\label{nufs2}
\end{eqnarray}
 appears in the sample-to-sample fluctuations at non-zero-energy density $e \ne 0$ (Eq. \ref{bce}),
while the 'true' correlation length exponent is nevertheless $\nu=1$ of Eq. \ref{nu1} in each sample, 
where the question is whether a given eigenstate of energy density $e$ is able to find resonances or not among the  $e^{L S_{\omega}(e)}$ other states of the same energy density $e$ 
(since states with a different energy density $e' \ne e$
cannot resonate by definition since the energy difference grows extensively as $E-E'=L(e-e')$. 
With respect to the general discussion (b) above, 
the very special feature of the MBL transition 
is that the exponent of Eq. \ref{nufs2} does not appear in the 
sample-to-sample fluctuations at zero-energy density $e = 0$ (see (i) above),
because in the middle of the spectrum, the number of available states grows exponentially as $2^L$
independently of the realization of the random fields.
In this model, the finite-size-scaling analysis based on various samples is thus much 'cleaner' 
at the middle of the spectrum $e=0$ than elsewhere in the spectrum $e \ne 0$.

\section{ MBL quantum spin chains with nearest-neighbor interactions } 

\label{sec_heisen}

\subsection{ Heisenberg chain with random fields }

The most studied model of MBL is 
the one-dimensional
Heisenberg chain with random fields $h_j$. The diagonal and off-diagonal parts of the Hamiltonian read respectively
\begin{eqnarray}
H_{diag} && =  \sum_{j=1}^L (h_j \sigma_j^z +J^{zz} \sigma_j^z \sigma_{j+1}^z )
\nonumber \\
H_{off} &&=  J \sum_{j=1}^L (\sigma_j^x \sigma_{j+1}^x+ \sigma_j^y \sigma_{j+1}^y) 
= 2 J \sum_{j=1}^L (\sigma_j^+ \sigma_{j+1}^-  + \sigma_j^- \sigma_{j+1}^+) 
\label{XXZh}
\end{eqnarray}
We consider that the disordered sample $\omega$ is fixed with its $L$ random fields $(h_1,...,h_L)$,
so that the control parameter is the coupling $J$ of the off-diagonal part.
Since the Hamiltonian conserves the total magnetization $\sum_i S_i^z$, let us focus on the zero magnetization sector :
the corresponding size $ {\cal N}_L$ of the Hilbert space is given by the number of configurations having $\frac{L}{2}$
positive and $ \frac{L}{2}$ negative spins
\begin{eqnarray}
{\cal N}_L = {L \choose \frac{L}{2}}  \opsimeq_{L \to +\infty}  \sqrt{ \frac{ 2 }{ \pi  L}  } 2^L
\label{zeromagn}
\end{eqnarray}

Here again, our goal is now to define a finite-size-pseudo-critical point $J_c(n,\omega,L)$
for each eigenstate $n$ of the sample $\omega$ of length $L$.

\subsection{ Forward Approximation for the eigenstates }

The Forward Approximation described in section \ref{sec_cayley} for the Anderson Localization on the Cayley tree \cite{alt_levitov,luca,forward}
has been extended to various MBL models \cite{forward,emergent_ros,qrem}.

For $J=0$, the off-diagonal part of the Hamiltonian of Eq. \ref{XXZh} vanishes.
So the eigenstates are simply given by the tensor products
\begin{eqnarray}
\vert \psi^{(0)}_{S_1,..,S_{L}} > && \equiv \vert S_1 > \otimes \vert S_2 > ... \otimes \vert S_{L} >
\label{hpsizero}
\end{eqnarray}
with the eigenvalues
\begin{eqnarray}
E^{(0)}_{S_1,..,S_{L}} =\sum_{j=1}^L (h_j S_j +J^{zz} S_j S_{j+1} )
\label{he0}
\end{eqnarray}

At first order in the coupling $J$, the eigenstates become
\begin{eqnarray}
 \vert \psi^{(0+1)}_{S_1,..,S_{L}} > 
&& =  \vert \psi^{(0)}_{S_1,..,S_{L}} >
+ \sum_{ \{S_i' \} } \vert \psi^{(0)}_{S_1',..,S_{L}'} > \frac{ < \psi^{(0)}_{S_1',..,S_{L}'} \vert H_{off} \vert \psi^{(0)}_{S_1,..,S_{L}} > }{ E^{(0)}_{S_1,..,S_{L}} -  E^{(0)}_{S_1',..,S_{L}'}}
\nonumber \\
&& =  \vert \psi^{(0)}_{S_1,..,S_{L}} >
+2 J \sum_{j=1}^L \sum_{ \{S_i' \} } \vert \psi^{(0)}_{S_1',..,S_{L}'} > \frac{ < \psi^{(0)}_{S_1',..,S_{L}'} \vert  (\sigma_j^+ \sigma_{j+1}^-  + \sigma_j^- \sigma_{j+1}^+)  \vert \psi^{(0)}_{S_1,..,S_{L}} > }{ E^{(0)}_{S_1,..,S_{L}} -  E^{(0)}_{S_1',..,S_{L}'}}
\label{heigenzeroper}
\end{eqnarray}
The possible local excitations are located on the antiferromagnetic bonds $S_j=-S_{j+1}$
and the corresponding energy difference reads
\begin{eqnarray}
  E^{(0)}_{S_1,..,S_{L}} 
- E^{(0)}_{S_1,.  -S_j,-S_{j+1} .,S_{L}} =2 S_j (h_j+ J^{zz} S_{j-1}) +2 S_{j+1}( h_{j+1}+J^{zz}S_{j+2} )
\label{diff1}
\end{eqnarray}
In the zero-magnetization sector that we consider, their number is typically of order
\begin{eqnarray}
N_{AF} \equiv  \sum_{j=1}^L (\delta_{S_j=-S_{j+1}}) \simeq  \frac{L}{2}
\label{naf}
\end{eqnarray}

To simplify the notations , let us now 
focus on some particular eigenstate
\begin{eqnarray}
 \vert 0 > && \equiv  \vert S_1,S_2... , S_{L} >
\label{heigenzero}
\end{eqnarray}
et denote the other states by the spin flips with respect to this state
\begin{eqnarray}
 \vert j,j+1 > && \equiv  \vert S_1,...,-S_j,-S_{j+1},..S_L >
\label{heigenj}
\end{eqnarray}
Then Eq. \ref{eigenzeroper} becomes
\begin{eqnarray}
 \vert \psi^{(0+1)}_{S_1,..,S_{L}} > 
&& =  \vert 0 >+ \sum_{j=1}^L (\delta_{S_j=-S_{j+1}}) \frac{J}{\epsilon_j} \vert j,j+1 > 
\label{heigenzeroperj}
\end{eqnarray}
with the notation (using $S_{j+1}=-S_j$)
\begin{eqnarray}
 \epsilon_j && \equiv  S_j (h_j+ J^{zz} S_{j-1}) + S_{j+1}( h_{j+1}+J^{zz}S_{j+2} )
\nonumber \\
&& =  S_j (h_j-h_{j+1}+ J^{zz} (S_{j-1}-S_{j+2} ))
\label{epsj}
\end{eqnarray}

At second order in the coupling $J$, if the locations $j_1$ and $j_2$ are not close $j_1+3<j_2$, the two energies $\epsilon_{j_1}  $ and $ \epsilon_{j_2} $ are independent, so that the amplitude of the eigenvector on $\vert j_1,j_1+1,j_2,j_2+1>$ 
reads by summing over the two possibles paths involving the two possible orders
\begin{eqnarray}
< j_1,j_1+1,j_2,j_2+1 \vert \psi^{(0+1)}_{S_1,..,S_{L}} > =  \frac{J}{\epsilon_{j_1} } \times \frac{J}{\epsilon_{j_1} +\epsilon_{j_2} } + \frac{J}{\epsilon_{j_2} } \times \frac{J}{\epsilon_{j_1} +\epsilon_{j_2} } = \frac{J^2}{\epsilon_{j_1} \epsilon_{j_2}}
\label{ampli2}
\end{eqnarray}
i.e. the amplitude reduces to the product of the two factors $\frac{J}{\epsilon_j}$ corresponding to the two local 
independent excitations, which is natural from a physical point of view.

More generally, at order $J^p$ with $p=\alpha \frac{L}{2}$, one may reach configurations 
displaying $2p=\alpha L$ spin flips with respect to the initial configuration.
If the density $\alpha$ is small $\alpha \ll 1$, these spin flips located at $(j_1,..,j_p)$
are diluted, and the corresponding amplitude
reduces to the product over $p$ independent factors as in Eq. \ref{ampli2}
\begin{eqnarray}
{\cal A}_{p=\alpha \frac{L}{2} } (j_1,j_1+1,j_2,j_2+1,..,j_p,j_p+1)  \opsimeq_{\alpha \ll 1} \prod_{k=1}^{p}  \frac{J}{\epsilon_{j_k} }
\label{amplin}
\end{eqnarray}
where the $\epsilon_j$ (Eq. \ref{epsj}) have for probability distribution
(the three spins $(S_j,S_{j-1},S_{j+2})$ are drawn with probabilities $p_{\pm}=\frac{1 }{2}$)
\begin{eqnarray}
P(\epsilon) && = \int dh_1 p(h_1)  \int dh_2 p(h_2)    
\left[  \frac{1}{2} \delta(  \epsilon -   h_1+h_2)
+ \frac{1}{4} \delta( \epsilon -   h_1+h_2+2 J^{zz}  )
+ \frac{1}{4} \delta( \epsilon -   h_1+h_2-2 J^{zz}  )
\right]
\label{pepsjz}
\end{eqnarray}

Of course when the density $\alpha$ of spin flips is not small, one should take into account the non-independence
of local excitations and an exact analysis of all cases becomes complicated.
As a consequence, it is useful to consider two opposite very simple approximations
to see the consequences on the scaling of pseudo-critical points.

\subsection{ Analysis based on typical properties of amplitudes }

As a first simple approximation, let us consider that the amplitude of the completely flipped configuration
can still be approximated by a product over $n \simeq \frac{L}{2}$ independent terms as in Eq. \ref{amplin}
\begin{eqnarray}
{\cal A} (1,2,...,L)  \simeq \prod_{k=1}^{\frac{L}{2}}  \frac{J}{\epsilon_k }
\label{amplifull}
\end{eqnarray}
so that its logarithm satisfies the Central Limit Theorem
\begin{eqnarray}
\ln \vert {\cal A} (1,2,...,L) \vert \simeq \sum_{k=1}^{\frac{L}{2}}  \ln \frac{J}{\vert \epsilon_k \vert }
\simeq  \frac{L}{2} \left[  \ln J - \overline{ \ln \vert \epsilon \vert }  \right] + x \sqrt{ \frac{L}{2} Var[ \ln \vert \epsilon \vert ] }  
\label{logamplin}
\end{eqnarray}
where $x$ is an $O(1)$ Gaussian variable.
As a consequence, if one defines the pseudo-critical point $J_c(n,\omega,L)$ as the coupling where the amplitude of Eq. \ref{amplifull} becomes unity, one obtains the scaling
\begin{eqnarray}
  \ln J_c(n,\omega,L) = \overline{ \ln \vert \epsilon \vert }   - x \sqrt{ \frac{2}{L} Var[ \ln \vert \epsilon \vert ] }  
\label{amplinpseudo}
\end{eqnarray}
involving the finite-size-scaling exponent 
\begin{eqnarray}
\nu_{FS}=2
\label{amplinnufs}
\end{eqnarray}
around the thermodynamic location of the critical point
\begin{eqnarray}
  \ln J_c(L=\infty) = \overline{ \ln \vert \epsilon \vert }  
\label{amplinjc}
\end{eqnarray}
In the localized phase $J< J_c$, the amplitude decays exponentially as
\begin{eqnarray}
\ln \vert {\cal A} (1,2,...,L) \vert 
\simeq  - \frac{L}{2} \left[  \ln \frac{J_c}{J} \right] \oppropto_{J \to J_c^-} - \frac{L}{2} \left[  \frac{J_c}{J} -1 \right]
\label{amplinloc}
\end{eqnarray}
involving the correlation length exponent
\begin{eqnarray}
\nu_{loc} =1
\label{nuloc1ampli}
\end{eqnarray}
These results are thus reminiscent of the scaling properties of pseudo-critical points
at Infinite Disorder Fixed Points \cite{us_tfic}.

\subsection{  Analysis based on the large deviation properties of amplitudes }

Let us now discuss the opposite approximation, where 
 the $ {L \choose \frac{L}{2}}   $ amplitudes involving $n=\frac{L}{2}$ denominators
\begin{eqnarray}
B_{n=\frac{L}{2}} = \prod_{j=1}^n  \frac{1}{\vert \epsilon_j \vert }
\label{bn}
\end{eqnarray}
are considered as independent.
Then the pseudo-critical point $J_c(n,\omega,L)$ 
 defined as the coupling where the maximum amplitude reaches unity
\begin{eqnarray}
1=\left[ J_c(0,\omega,L) \right]^{\frac{L}{2}}   \opmax_{1 \leq p \leq {L \choose \frac{L}{2}} }   \left(  B_{\frac{L}{2}}(p) \right)
\label{hmaxtree}
\end{eqnarray}
will depend on the large-deviation properties of the logarithms of the amplitudes
\begin{eqnarray}
Y_n =\ln B_n =  \sum_{j=1}^n  \ln \frac{1}{\vert \epsilon_j \vert }
\label{logbn}
\end{eqnarray}
so that the discussion is very similar to the analysis of section \ref{sec_cayley} concerning the Anderson Localization on the Cayley tree.
The exponentially small probability in $n$ to see an anomalously large amplitude $Y_n \simeq n y$ is governed by some rate function $I(y)$ 
\begin{eqnarray}
{\cal P}_n \left( y=\frac{Y_n}{n} \right) \propto e^{-n I(y)}
\label{bnlargedev}
\end{eqnarray}

The probability distribution $Q_{max}(y_{max}) $
of the maximum $y_{max}$ among ${L \choose \frac{L}{2}}$ such variables $y$
can be obtained from the cumulative distribution
\begin{eqnarray}
\int^y dy_{max} Q_{max}(y_{max}) && = \left[ 1- \int_{y}^{+\infty} dy'  {\cal P}_{\frac{L}{2}} (y' ) \right]^{{L \choose \frac{L}{2}}}
\simeq e^{- {L \choose \frac{L}{2}} \int_{y}^{+\infty} dy'  {\cal P}_{\frac{L}{2}}(y' ) } \simeq e^{- 
\sqrt{ \frac{ 2 }{ \pi  L}  } 2^L \int_{y}^{+\infty} dy'  e^{- \frac{L}{2} I(y') } }
\nonumber \\
&& \simeq  e^{- \sqrt{ \frac{ 2 }{ \pi  L}  } e^{\frac{L}{2}\left[ 2 \ln 2 -  I(y) \right] }}
\label{hcumulqmax}
\end{eqnarray}

It is thus useful to introduce the value $y^*_L$ satisfying
\begin{eqnarray}
 1=\sqrt{ \frac{ 2 }{ \pi  L}  }  e^{ L \left[ \ln 2-  \frac{I(y^*_L) }{2} \right] } 
\label{ystarl}
\end{eqnarray}
and
\begin{eqnarray}
b_L \equiv \frac{2}{L I'(y^*_L)} 
\label{blstarl}
\end{eqnarray}
Then the change of variables $y=y^*_L+b_L x$ in Eq. \ref{cumulqmax}
yields the convergence towards the Gumbel distribution for the $O(1)$ variable $x$
\begin{eqnarray}
\int^{y^*_L+b_L \xi} dy_{max} Q_{max}(y_{max}) 
&& \simeq   e^{-   e^{ -  x  } } 
\label{hcumulqmaxxi}
\end{eqnarray}

The pseudo-critical-point of Eq. \ref{maxtreea} becomes
\begin{eqnarray}
 J_c(0,\omega,L)
&& 
 = \left[  e^{\frac{L}{2} y_{max}}  \right]^{-\frac{2}{L}} \simeq e^{-y_{max}}  = e^{-y^*_L-b_L x}
\label{jcdistri}
\end{eqnarray}
where $x$ is an $O(1)$ random variable drawn with the Gumbel distribution (Eq. \ref{cumulqmaxxi}). 
In the thermodynamical limit $L \to +\infty$, the location of the transition 
is given by
\begin{eqnarray}
J_c(L \to +\infty) = e^{-y^*_{\infty} }
\label{jcthermo}
\end{eqnarray}
where $y^*_{\infty}$ is the solution of
\begin{eqnarray}
I(y^*_{\infty}) =  2 \ln 2 
\label{iystar}
\end{eqnarray}
The solution $y^*_L$ of Eq. \ref{ystarl} reads for large $L$
\begin{eqnarray}
y_L^* \simeq y^*_{\infty} - \frac{2}{L  I'(y^*_{\infty})}  \ln \left( \sqrt{ \frac{ \pi  L}{2}  } \right) 
\label{ystarlsolu}
\end{eqnarray}

At leading order for large $L$, the pseudo-critical point of Eq. \ref{jcdistri}
displays the scaling
\begin{eqnarray}
 J_c(0,\omega,L) && = J_c(L=\infty)   e^{-(y^*_L-y^*_{\infty} )-b_L x} 
 \opsimeq_{L \to +\infty}   J_c(L=\infty)
 \left[  1 +\frac{2}{L  I'(y^*_{\infty})}  \ln \left( \sqrt{ \frac{ \pi  L}{2}  } \right)  - \frac{2}{L I'(y^*_{\infty})}  x \right]  
\label{jcdistrifinal}
\end{eqnarray}

The convergence in $ \frac{1}{L}$ corresponds to the correlation length exponent 
\begin{eqnarray}
\nu=1
\label{heisennu}
\end{eqnarray}
in agreement with the numerical studies giving $\nu \simeq 0.8 (3)$ \cite{kjall,alet}.

\section{ Conclusion } 

\label{sec_conclusion}

To better understand the finite-size-scaling properties of the Many-Body-Localization transition,
we have proposed to associate to each eigenstate a finite-size-pseudo-critical point and to study its statistical properties governed by the correlation length exponent $\nu$. We have first explained how this idea works for the case of the Anderson Localization Transition concerning a single particle,
both in the presence of long-ranged hopping or in the presence of nearest-neighbor hoppings on the Cayley tree.
We have then studied in detail the MBL quantum spin chain toy model of Ref \cite{c_mblper},
where the eigenstate-to-eigenstate fluctuations involve the exponent $\nu=1$, while the sample-to-sample 
fluctuations introduce additional fluctuations involving $\nu_{FS}=2$ but only at non-zero-energy density $e \ne 0$.
Finally for the Heisenberg chain in random fields, we have proposed  
two simple approximations for the scaling of pseudo-critical points, that both correspond to $\nu_{loc}=1$,
while the finite-size exponent can be either $\nu_{FS}=2$ or $\nu_{FS}=1$.

All the models that we have considered have been analyzed via some strong disorder perturbative expansion.
Further work is needed to see how this idea of pseudo-critical points can be used numerically.
Various criteria can be used, and
it is actually a good idea to
consider various definitions and to compare them, as for other quantum transitions concerning ground-states of random quantum spin chains where for instance three different criteria have been compared \cite{us_tfic}.
Besides the exponent $\nu$, 
this alternative analysis of finite-size-scaling 
 usually allows to better measure the other critical exponents
 characterizing the critical state itself : 
 for instance for the quantum Ising model in various dimensions,
this method has been very useful
 to measure the fractal dimension of the critical cluster
\cite{kovacsstrip,kovacs2d,kovacs3d,kovacsreview,kovacsLR,kovacsLR3d}.
For the present MBL transition, 
one can similarly expect that the properties of the critical
eigenstates will be much clearer if one considers the behavior of each eigenstate at its finite-size-pseudo-critical point, where it is 'truly critical',
instead of using standard averages over eigenstates and samples that see a mixture of states that are effectively either localized or delocalized. This point is especially important for the entanglement entropy, which is one of the most studied order parameter of the MBL transition. The recent numerical study \cite{luitz_bimodal} has reported the bimodal mixture of volume-law eigenstates and of area-law eigenstates even within a given disordered sample at a given energy density, so that the average yields a volume-law with a reduced non-thermal coefficient, although no individual state displays this behavior. So this averaging seems to hide the growth behavior of the entanglement entropy of 'true critical eigenstates'. In conclusion, even if this analysis in terms of finite-size-pseudo-critical points for individual eigenstates is much more complicated in practice than the standard finite-size-scaling, we feel that it is essential to identify true critical eigenstates and hence to clarify the nature of the Many-Body-Localization transition.

\section*{ Acknowledgments }

It is a pleasure to thank David Luitz, Fabien Alet and Nicolas Laflorencie for 
very useful mail discussions about the results of the study \cite{luitz_bimodal} 
that have been at the origin of the present work.

\end{document}